\documentclass[a4paper,twoside]{article}

\usepackage{epsfig}
\usepackage{subfigure}
\usepackage{calc}
\usepackage{amssymb}
\usepackage{amstext}
\usepackage{amsmath}
\usepackage{amsthm}
\usepackage{multicol}
\usepackage{pslatex}
\usepackage{apalike}
\usepackage{SciTePress}
\usepackage[small]{caption}
\usepackage{graphicx}

\begin{document}

\title{\uppercase{An analysis of social network connect services}\thanks{Preprint of article published in Proceedings of WEBIST 2012. Porto, Portugal}}

\author{\authorname{Antonio Tapiador\sup{1}, V\'{i}ctor  S\'{a}nchez\sup{1} and Joaqu\'{i}n Salvach\'{u}a\sup{1}}
\affiliation{\sup{1}Universidad Polit\'{e}cnica de Madrid, Av. Complutense 30, Madrid, Spain}
%\affiliation{\sup{2}Department of Computing, Main University, MySecondTown, MyCountry}
\email{\{atapiador, vsanchez, jsalvachua\}@dit.upm.es}
}

\keywords{
%The paper must have at least one keyword. The text must be set to 9-point font size and without the use of bold or italic font style. For more than one keyword, please use a colon as a separator. Keywords must be titlecased.
Social Networks; API; OAuth; REST: mashups
}

\abstract{
% One sentence on motive, method, key results, conclusions
Social network platforms are increasingly becoming identity providers and a media for showing multiple types of activity from third-party web sites. In this article, we analyze the services provided by seven of the most popular social network platforms. Results show OAuth emerging as the authentication and authorization protocol, giving support to three types of APIs, client-side or Javascript, server-side or representational state transfer (REST) and streaming. JSON is the most popular format, but there a considerable variety of resource types and a lack of representation standard, which makes harder for the third-party developer integrating with several services.
}

\onecolumn \maketitle \normalsize \vfill

\section{\uppercase{Introduction}}
\label{sec:introduction}

% Start with a good first sentence!

%% What is the problem and why is it interesting?

\noindent Popular social network platforms are leveraging identity services. They provide authentication commodities, relieving third-party websites, applications and services from managing user data. They are also an increasingly important media. Using already existent identity providers can improve the engagement of users in those third-party sites. Through APIs, these providers allow third-party applications to connect with the activity streams of the users and insert new stories created by the use of the third-party application or service. Nowadays it is not strange to see brands, companies and products offering strong integration with Facebook, Twitter and others providers to improve engagement, visibility and impact.

%% Who are the main contributors?
%% What did they do?

Web APIs have been used in last years to create mashups, a composition of services that facilitate the design and development of modern Web applications \cite{Benslimane:2008}. Nowadays, social network platforms are between the most popular Web sites \cite{Mislove:2007}, though their history is very recent \cite{Boyd:2007}.

Ko et al. review social network connect services from Facebook, Google and Myspace \cite{Ko:2010}. They show how these services provide social network features to third-party websites, which do not have to build their own social network. They easy sign in and enrich user data and experience by mashing up their own data with the pieces retrieved from the API for instance: finding friends in the platform. On the other hand, this is also interesting for the provider that is now showing new kinds of activity in the streams.

An attempt to provide a unified connect service was made by OpenSocial \cite{Hasel:2011}, a standard promoted by Google, Myspace and others, which has been adopted to some extend.

%% What novel thing will you reveal?

In this article, we provide an in deep analysis of current social network connect services. We have reviewed seven popular social network platforms. First of all, there is an analysis of OAuth, the protocol used by all of them as sign in and authorization technology. The OAuth protocol has several versions, modes, permissions that every provider we have analyzed supports in a different degree. Besides there is an analysis of the available APIs featuring similarities and differences between them. Several patters have been identified, i.e. client oriented or Javascript, server oriented or REST and real-time/streaming oriented. We enumerate the widget types used in Javascript APIs. For REST APIs, we analyze the formats and resource representations offered by the social network connect services. Finally, we analyze the streaming APIs.

\section{\uppercase{Method}}

%% Experimental paper: equipment, materials, method
%% Modeling paper: assumptions, mathematical tools, method
%% Computational paper: inputs, computational tools, method

%% Explain what is especially different about your method
%% Give sufficient detail that the reader can reproduce what you did

We have reviewed the APIs of seven popular social networking services; Facebook\footnote{http://www.facebook.com/}, the popular and well known social network platform; Twitter\footnote{http://twitter.com/}, the popular microblogging platform; Google+\footnote{http://plus.google.com/}, the recent Google's bet in social networks; LinkedIn\footnote{http://www.linkedin.com/}, the social network platform for professional contacts; Github\footnote{https://github.com/}, the most popular platform for developers and social code sharing; Myspace\footnote{http://www.myspace.com/}, the former popular champion in social networking and Foursquare\footnote{https://foursquare.com/}, the places check-in oriented social network.

These platforms offer extensive documentation on their APIs as well as online tools for testing them.

\subsection{OAuth}

OAuth is the omnipresent \textit{"open protocol to allow secure API authorization in a simple and standard method from desktop web applications"} \footnote{http//oauth.net/}. It was developed by some web enthusiastics in the aim to develop a common standard for accessing APIs. Version 1.0 was proposed as a IETF request for comments \cite{rfc5849}.

OAuth removes an anti-pattern that was appearing in web services applications. With the grown of those applications, there was a need of a web application to access protected resources from others. For instance, imagine some private pictures in a photo sharing service, and another web application that prints photos and send them to your home. You may want to let the printing service access to several private photos in order to print them for you. Before OAuth, the only way to achieve this was sharing the photo-sharing web service credentials to the printing service. The obvious drawback of this approach was that the printing service had full access to your account in that photo-sharing service.

OAuth proposes a solution to this problem. When the user wants to print the photos, the photo printing service will redirect the user-agent to the OAuth handler of the photo-sharing service. In this step the user will authorize the photo-sharing service to provide to the third-party app (printing service) access to their pictures. Finally the request will be redirected to the printing service including an access token parameter. Using this provided token the printing service will be granted access to the user private photos.

In April 29, 2009 a security flaw was discovered in the protocol \footnote{http://oauth.net/advisories/2009-1/}. A session fixation attack allowed an attacker to pre-build an authorization request and inject it to a user to finish it. OAuth version 1.0a was released. This security announce was about a year before the RFC publication, so the IETF's document already has the security fix. Nevertheless, most of the APIs still refer to their supported version of OAuth as 1.0a.

OAuth 2.0 is the next version of the protocol. It is about to be published as RFC by the IETF. OAuth 2.0 simplifies the protocol. It describes four different authorization flows, i.e., \textit{authentication code}, \textit{implicit}, \textit{resource owner password credentials} and \textit{client credentials}. The first one, authorization code, is suitable for clients capable of maintaining the confidentiality of their credentials, i.e. web applications that reside in a web server. This flow authenticates the client, and the authorization token is granted directly to the client, it does not pass through the user-agent. The second one, implicit flow, is suitable for clients implemented in a browser, typically Javascript. The access token is directly granted to the client, it is not authenticated neither. The third is similar to the case OAuth tries to solve. The client uses the user's credentials to take an authorization token though they are used only once and are not stored. Nevertheless, this method is requested to be the last alternative to be used. Finally, client credentials allows a client to manage protected resources controlled directly by him in the authentication server.

OAuth 2.0 also describes access token scopes. The scope parameter contains a list of space-delimited parameters defined by the authentication server. They are used to define to which resources the user is granting access to the third-party application or service.

\subsection{API types}

Social web platforms support several types of APIs; client side or Javascript, web server side or representational state transfer (REST) and streaming APIs.

\subsubsection{Client side or Javascript API}

The most straightforward way to integrate third-party web sites with social network platforms is through Javascript APIs. Social platforms provide Javascript libraries that can be included and used in any web site. It is an easy way to mashup local content with content from the social platform. There are a lot of applications that go from the popular "like" buttons or authentication to product recommendations based on social data. This APIs use OAuth's implicit flow. All Javascript APIs use JSON format due to the fact that does not need any really parsing step.

\subsubsection{Web server side or REST API}

% All the platforms implement a REST-oriented APIs, which are more soft and easy to use. Hablar un poco del giro hacia web services y entroncarlo con la antigua guerra contra los Web Services (soap). Quizá en la intro?

Web-server-side APIs are oriented to server-based or desktop-based clients. They just provide data without any view-related content. They all follow the REST principles of addressability, stateleness and representations \cite{Fielding:2000,Richardson:2007}. REST principles and REST-oriented architectures encourage the use of standard HTTP verbs \cite{rfc2616,rfc5789}; such as \texttt{GET} for reading, \texttt{POST} for adding resources, \texttt{PUT} for replacing a given resource, \texttt{PATCH} for changing an existing resource and finally \texttt{DELETE} for removing resources. REST APIs also support a variety of representations. Most common formats include Javascript's format JSON, XML, and their syndication extensions, RSS and ATOM. The resource representations REST API can also provide some commodities or extensions, such as pagination for large collections of data, or introspection that allows the examination of the resource's metadata. They provide each resource with a URL, so their representation can be retrieved and they can also be referenced from other resources.

The resources' addresses provided by REST APIs can be of different kinds. There can be a single resource URL scheme, or divided by resource type. In the case of a single URL scheme, the resource's representation provides a field declaring the kind of resource.

As is expected, every social network has a resource type to represent the user. Another metric we have used is comparing these fields to see if they follow any standard, which could be interesting to provide a uniform interface to the third-party developer that uses two or more social network providers.

% Privacy deepness in the API. Separation degree. %%(TO-DO, if not DELETE THIS LINE!!!)

\subsubsection{Real time or Streaming API}

The other kind of API provided by social network platforms is the streaming API. Unlike the former two APIs, in which the client request the platform for data, this goes the other way. The client subscribes to the platform, which pushes data to the client. This API is suitable for real time applications. 

\section{\uppercase{Results}}

%% Present the output of the experiments, model or computation

\begin{figure}
\centering
\includegraphics[scale=0.1]{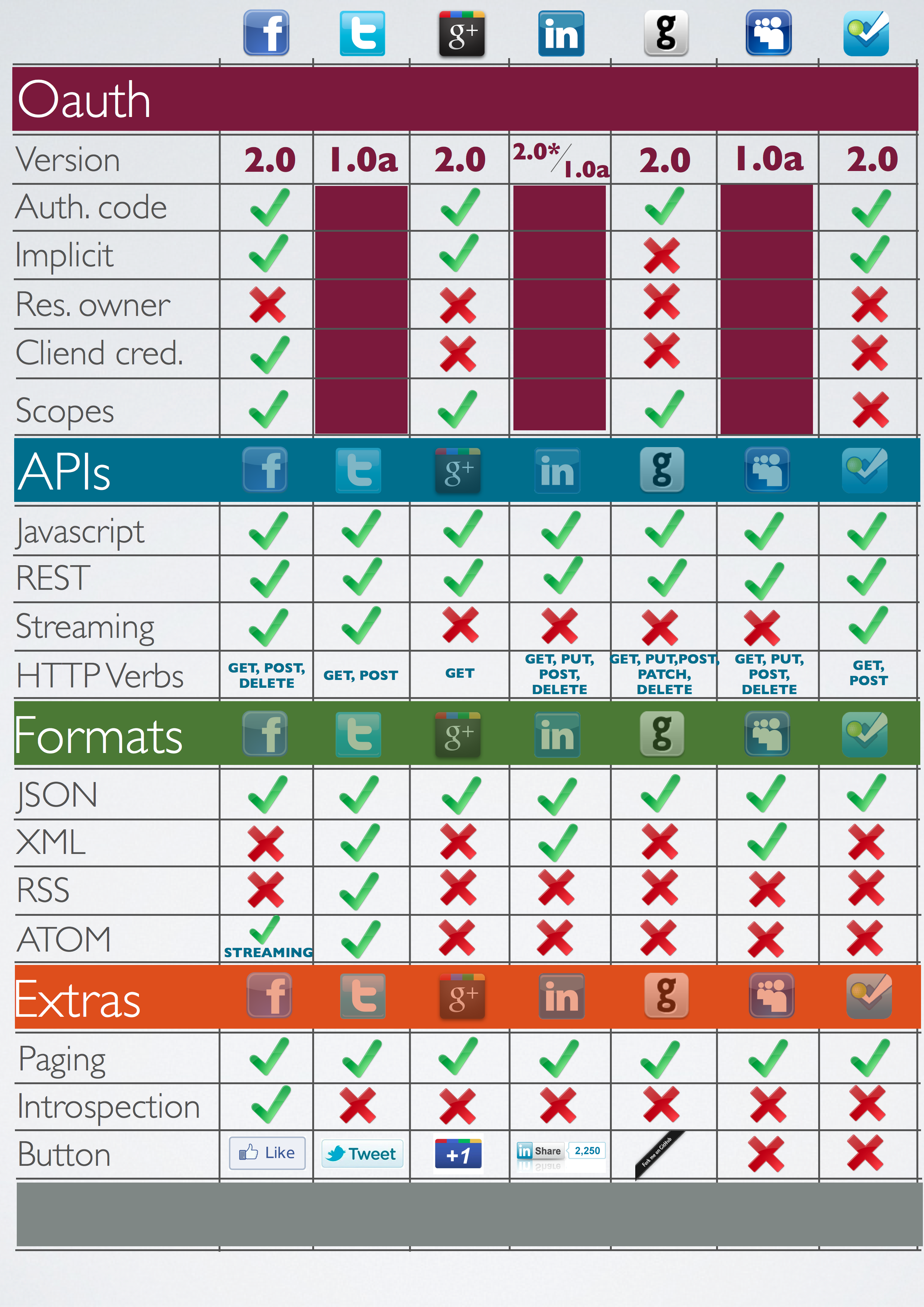}
\caption{Analysis of the connect services of seven of the most popular social network platforms}
\label{fig:results}
\end{figure}

On this section, we present the result of the analysis of the seven social network platforms mentioned above. Figure \ref{fig:results} shows a summary of these results.

\subsection{OAuth}

While all the platforms support OAuth, the versions are currently divided between 1.0a (Twitter, LinkedIn and Myspace), and 2.0 (Facebook, Google+, Github and Foursquare). Claims from users can be found in developer forums asking providers to support version 2.0, so it is probable that version 2.0 is implemented by all the platforms in the short term.

Between the platforms providing version 2.0, the support for authentication flows is disparate. All of them implement the web server/authentication code flow, giving support to web applications. The next one supported is the implicit flow, primary used in the context of Javascript applications. This is implemented by three of four platforms. Even though LinkedIn claims to be using OAuth 2.0 in their Javascript API, their implementation is not standard, so it is not considered as that. Finally, client credentials flow is only supported by Facebook. They use it to provide third-party applications access to pending requests (\texttt{apprequests}) in the Facebook platform. They are the clear champion in OAuth 2.0 implementations.

Regarding OAuth scopes, their definition is very disparate and significant. Facebook defines more than 60 different scope types, such as \texttt{user\_about\_me}, \texttt{friends\_about\_me}, \texttt{user\_activities}, \texttt{friends\_activities}, \texttt{user\_birthday}, \texttt{friends\_birthday}, etc. Most of them are classified in information from a user and information from his friends. Many of the scopes are related to the profile items, many others with collection items types like pictures, videos, etc. The special scope \texttt{manage\_pages} provides an access token for impersonating pages. It is way of using the API impersonating the page. A complete reference on permissions can be found online\footnote{http://developers.facebook.com/docs/reference/api/permissions}. On their side, Google+ has a full variety of OAuth scopes\footnote{http://code.google.com/apis/gdata/faq.html\#AuthScopes}. They are providing a different namespace of permissions for each Google application (Analytics, Blogger, Calendar, Gmail, etc.), so their architecture is more modular than Facebook's. Github support only 4 types of scopes\footnote{http://developer.github.com/v3/oauth/\#scopes} (\texttt{user}, \texttt{public\_repo}, \texttt{repo}, \texttt{gist}), which are very focused on their application area: social coding. Finally, Foursquare do not use OAuth scopes.

\subsection{APIs}

\subsubsection{Client side or Javascript API}

Almost all the platforms provide a login button. This feature allows the third-party application delegating authentication and profile information to the social network. It is implemented by Facebook, Twitter, Google+, Github and LinkedIn. In the case of Google+, the login plug-in comes from Google accounts; this authentication service is not exclusive from Google+ and is used in every single Google product.

Another popular functionality is the "like" button, that allows users to post one site to their social network at the same time they are supporting it, since they are working towards a model where more likes means more quality. This is supported by Facebook (\textit{like}), Google+ (\textit{+1}) and LinkedIn (\textit{Share this in LinkedIn}). There are other buttons available. Facebook has the bigger offer, including the send button to post to the Facebook wall, comments plug-in to comment about an activity, activity feeds and recommendations, as well as the login and register buttons. Twitter has the "\textit{tweet this}", to publish a new tweet about the content of the third-party web site, or the tweet content that displays the number of tweets about the page. Github provides a "\textit{fork me on github}" button that points to the open source repository.

\subsubsection{Web server side or REST API}

\paragraph{HTTP verbs}

Though the REST principles suggest using an HTTP per action, this issue is followed to different degree by current social network platforms. In one side it is Google+, that supports only read-only operations through \texttt{GET}. We expect that they support write operations soon. The next group of APIs include Twitter and Foursquare. These platforms only use \texttt{GET} and \texttt{POST}. They both support updating and deleting operations, but they are performed using \texttt{POST}. Facebook follows, using the \texttt{DELETE} verb. Updating is rare in Facebook, but it is supported via \texttt{POST}. LinkedIn and Myspace perform these operations with a more appropriate HTTP verb: \texttt{PUT}, which was designed for operations updating resources. Finally, Github is the champion of this category, introducing yet another HTTP verb: \texttt{PATCH}. It is used for updating only selected fields from a resource, instead of \texttt{PUT} which is intended to update the whole resource.

\paragraph{Resource representations}

JSON is the omnipresent representation format, supported by every analyzed social network platform. It fits very well with Javascript APIs. Some of the platforms, such as Facebook, Google+, Github and Foursquare only support this format. Mostly all the platforms also support JSONP, allowing the API users to add a callback to JSON calls and avoid restrictions related to same origin policies \cite{Oehlman:2011}. XML is supported as second format by LinkedIn and Myspace. Finally, Twitter is the champion in this category, supporting also XML derived formats such as RSS and ATOM.

% Facebook (25) Achievement(Instance), Album, Application, Checkin, Comment, Domain, Event, FriendList, Group, Insights, Link, Message, Note, Order, Page, Photo, Post, Question, QuestionOption, Review, Status message, Subscription, Thread, User, Video
% Twitter (13) Tweets, Direct message, Friends, User, Favourites, Lists, Account, Notifications, Saved searches, Trends, Geo, Blocks, Spam 
% Google+ (3)  people, activities and comments.
% LinkedIn (8) people, companies, jobs, mailbox, network/updates, shares, comments, likes
% Github (25) gist, comment, blobs, commits, references, tags, trees, issues, labels, milestones, orgs, members, teams, pulls, repos, collaborators, downloads, forks, keys, watching, hooks, users, emails, followers, keys
% Foursquare (12) users, venues, venuegroups, checkins, tips, lists, updates, photos, settings, specials, campaings, events
% Myspace (10) activities, albums, app_data, groups, media_items, notifications, people, profile_comments, status_mood, stream_subscription

The resource types supported by each network vary from 3 by Google+ to more than 25 by Facebook. The are very different and depend on the focus of the social network. For instance, Github has \textit{repositories}, \textit{commits} and \textit{forks}, while Foursquare support \textit{venues} and \textit{checkins}.

\begin{figure*}
\centering
\includegraphics[scale=0.3]{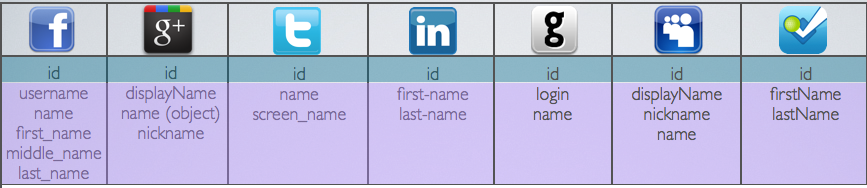}
\caption{Id and name related fields in the JSON representation of users. While the id field is a common parameter, there are not two social network services using the same fields for names.}
\label{fig:user}
\end{figure*}

Nevertheless, all of the social network platforms support the user resource, which represent the users in the platform. We have analyzed the JSON representation of these resources. The results show that there is only one parameter in common: the \texttt{id}. All the other parameters are totally different in all networks. Figure \ref{fig:user} shows this issue. It gathers name-related parameters in each social networks. It is representative how there is not even two social network services using the same parameters for user names. The rest of resource fields are even more disparate.

% Searching?

% Pagination:
% Facebook  request: limit, offset, until, since
%           response: paging: previous, next
% Twitter: request: page, count
%          response: 
%
% Github request: page, per\_page
%        response: link header, rel= next, last, first, prev

Something similar happens with resource pagination. Every social network uses different parameters for requests (e.g. \texttt{limit}, \texttt{offset}, \texttt{page}, \texttt{count}, \texttt{per\_page}) and response (e.g. \texttt{previous}, \texttt{prev})

\paragraph{Resource addresses}

The common case in REST APIs is providing separate end-points for resources. Facebook is the only network that provides a single API address replaced their old-traditional REST API for the new Graph API\footnote{http://developers.facebook.com/docs/reference/api/}. It is a single interface to all the objects they manage. They also use the Opengraph protocol\footnote{http://developers.facebook.com/docs/opengraph/} for adding metadata to the HTML of web pages, using the \texttt{meta} tag in the HTML's \texttt{head}. This way you can describe metadata like \texttt{title}, \texttt{image}, \texttt{url}, but it is redundant information, since even the Facebook's parser is able to obtain this information from other means (standard HTML tags). An interesting feature of the Opengraph is that Facebook uses this data to connect with Facebook applications and give admin permissions to users, through a couple of specific tags \texttt{fb:admins} and \texttt{fb:app\_id}.

\subsubsection{Real time or Streaming API}

Real time APIs allow third-party applications subscribing to updates from the social network platform. The most real-time of the APIs is Twitter's. They allow to get opened an HTTP connection and receiving a streaming of JSON objects, without closing the connection.

On the other hand, Facebook, Foursquare and Myspace support their platform performing HTTP requests to a third-party endpoint when a new event must be notified. Facebook uses an standard here, pubsubhubub \footnote{http://code.google.com/p/pubsubhubbub/}, while Foursquare requires the consumer to register on their web and using HTTPS. Posts are performed using JSON.

\section{\uppercase{Discussion}}

%% Extract principles, relationships, generalizations
%% Present analysis, model or theory
%% Show relationship between the results and analysis, model or theory

All the social network connect services analyzed use OAuth at least in version 1.0a. The ones that use 2.0, they all implement web server support, and most of them client-side support. Authorization scopes are very particular to each API and there are not enough cases to obtain patterns. Nevertheless, we can observe two different models in architecture, Facebook's and Google's. While Facebook has a lot of different permissions, due to the centralized nature of their service, Google uses URI's in scopes, because of their decentralized architecture of services. This matter provides Google with better data isolation and more privacy. It is also interesting how they ask the user for permissions in both cases, for Google it seems to be enough asking for "Agenda permissions", Facebook on the other hand translates a list of scopes in blocks of user friendly text.

We also want to point to the fact that the change from other authentication solutions, like OpenID, to OAuth makes things harder to other identity services. OpenID uses URI identifiers, so all the services were at the same level. With OAuth, the third-party developer must put a sign in button for every identity provider he is willing to support. Nevertheless, this issue was already appearing in OpenID implementations, where there were special buttons for most popular providers besides the uniform OpenID identifier field.

Regarding Javascript APIs, they are the quickest way of integrating a third-party applications with a social network services. We can see authentication widgets being very popular, with the ability to import authentication, profile information and even contacts to a third-party. Other plug-ins that are also very popular are the ones providing the ability to post to one's stream, both in the simpler form of likes as well as a more personal posts. Google uses the information of their +1 button to improve searches on their main product (searcher) that now will show the results that our friends recommend as featured entries. The use of Google +1 button show us how the information generated by the users in social network can be used, and is currently being used to improve user experience, of course privacy needs to be respected and courts all over the world are trying to define clear boundaries.
Facebook is also the winner in this category with a wide range of plug-ins.

REST APIs are diverse and do not follow a clear standard, which makes things harder to the third-party developer that wants to integrate with several services, because he needs a specific library for all of them. The support for REST principles in HTTP verbs seems to be disparate, with some of them supporting them versus other of them overloading POST. The JSON format is clearly the preferred in web services. Resource representations are very disparate and their are all focused in their social network field. This is true even for the representation of a common resource: the user and also for the pagination extensions along all the social network services. We think a standard should emerge in this field, maybe the Activity Streams \footnote{http://activitystrea.ms/} standard could be an option where OpenSocial has not been.

In a very different direction we have Paul Kinlan's (Google Developer Advocate) webintents. Webintents idea \footnote{http://webintents.org/} revolves around the fact that APIs are further than ever to be standard but the need of using them is rapidly increasing. Webintents mimics Android intents where we "invoke" other apps to make an specific task, then the app returns the results that the original app uses to any purpose. If we could do the same on the internet there would not be a reason to rewrite thousands of already existing APIs. Webintents declare the services that our API provides and other apps can use by invoking us, just by knowing the API end-point.

Finally, it is worth mentioning how technology is evolving from request based APIs to push based APIs. Nowadays when we constantly update our Twitter or Facebook applications we are generating alongside the other millions of users an overwhelming amount of refresh requests that are not that different from a distributed denial of service (DDoS) attack. Due to this fact, developers and social networks implemented different techniques such as AJAX or long polling. But the optimal solution is creating a duplex channel between client and server, so it is easy to justify the appearance of technologies such as push notifications, like Twitter's streaming API shown above, and HTML5 websockets \footnote{http://websocket.org/}.

\section{\uppercase{Conclusion}}

%% Draw together the most important results and their consequences
%% List any reservations or limitations

Social network platforms are becoming identity providers and media for communication. Their services are increasingly being used by third-party web applications. For an analysis of this services, we can see OAuth as an emerging standard for authentication and authentication, giving support for client-side Javascript APIs and server-side REST APIs. Javascript plug-ins are increasingly popular and provide a way to delegate authentication, as well as promote third-party services in the social network platform. Server-side APIs follow REST principles, use JSON formats and do not follow any standard for resources representation. Streaming APIs are getting more popular to optimize information transmission.

\vfill
\bibliographystyle{apalike}
{\small
\bibliography{books,standards,rfc,social-networks,web-architecture}}

\begin{thebibliography}{}

\bibitem[Benslimane et~al., 2008]{Benslimane:2008}
Benslimane, D., Dustdar, S., and Sheth, A. (2008).
\newblock Services mashups: The new generation of web applications.
\newblock {\em Internet Computing, IEEE}, 12(5):13 --15.

\bibitem[boyd and Ellison, 2007]{Boyd:2007}
boyd, d.~m. and Ellison, N.~B. (2007).
\newblock Social network sites: Definition, history, and scholarship.
\newblock {\em Journal of Computer-Mediated Communication}, 13(1):210--230.

\bibitem[Dusseault and Snell, 2010]{rfc5789}
Dusseault, L. and Snell, J. (2010).
\newblock {PATCH Method for HTTP}.
\newblock RFC 5789 (Proposed Standard).

\bibitem[Fielding, 2000]{Fielding:2000}
Fielding, R. (2000).
\newblock Architectural styles and the design of network-based software
  architectures.
\newblock {\em Doctoral dissertation}.

\bibitem[Fielding et~al., 1999]{rfc2616}
Fielding, R., Gettys, J., Mogul, J., Frystyk, H., Masinter, L., Leach, P., and
  Berners-Lee, T. (1999).
\newblock {Hypertext Transfer Protocol -- HTTP/1.1}.
\newblock RFC 2616 (Draft Standard).
\newblock Updated by RFCs 2817, 5785, 6266.

\bibitem[Hammer-Lahav, 2010]{rfc5849}
Hammer-Lahav, E. (2010).
\newblock {The OAuth 1.0 Protocol}.
\newblock RFC 5849 (Informational).

\bibitem[H\"{a}sel, 2011]{Hasel:2011}
H\"{a}sel, M. (2011).
\newblock Opensocial: an enabler for social applications on the web.
\newblock {\em Commun. ACM}, 54:139--144.

\bibitem[Ko et~al., 2010]{Ko:2010}
Ko, M.~N., Cheek, G., Shehab, M., and Sandhu, R. (2010).
\newblock Social-networks connect services.
\newblock {\em Computer}, 43(8):37 --43.

\bibitem[Mislove et~al., 2007]{Mislove:2007}
Mislove, A., Marcon, M., Gummadi, K.~P., Druschel, P., and Bhattacharjee, B.
  (2007).
\newblock Measurement and analysis of online social networks.
\newblock In {\em Proceedings of the 7th ACM SIGCOMM conference on Internet
  measurement}, IMC '07, pages 29--42, New York, NY, USA. ACM.

\bibitem[Oehlman et~al., 2011]{Oehlman:2011}
Oehlman, D., Blanc, S., Oehlman, D., and Blanc, S. (2011).
\newblock Integrating with social apis.
\newblock In {\em Pro Android Web Apps}, pages 221--254. Apress.

\bibitem[Richardson and Ruby, 2007]{Richardson:2007}
Richardson, L. and Ruby, S. (2007).
\newblock {\em Restful web services}.
\newblock O'Reilly, first edition.

\end{thebibliography}

\vfill
\end{document}